\pdfoutput=1
\RequirePackage[OT1]{fontenc}
\documentclass[letterpaper, 10 pt, conference]{ieeeconf}
\IEEEoverridecommandlockouts
\usepackage{amssymb, amsfonts, mathtools, xargs, tensor, units, cite, stmaryrd, mathrsfs, algpseudocode, algorithm, graphicx,subcaption,tikz,pgfplots, xcolor,cancel,multicol} 
\usepackage{tikz}
\usetikzlibrary{3d, calc}
\usepackage{pgfplots}
\usepgfplotslibrary{groupplots}
\usepackage{pgfplotstable}
\usepackage{booktabs}
\pgfplotsset{compat=1.18}
\usepackage[unicode=true, colorlinks=false, hidelinks]{hyperref}
\usepackage{comment}
\newtheorem{assumption}{Assumption}

\newtheorem{remark}{Remark}

\definecolor{color1}{RGB}{0,128,0}      %
\definecolor{color2}{RGB}{255,0,0}      %
\definecolor{color3}{RGB}{0,0,255}      %
\definecolor{color4}{RGB}{0,0,0}        %
\definecolor{color5}{RGB}{255,140,0}    %
\definecolor{color6}{RGB}{75,0,130}     %
\definecolor{color7}{RGB}{220,20,60}    %
\definecolor{color8}{RGB}{128,128,0}    %
\definecolor{color9}{RGB}{0,139,139}    %
\definecolor{color10}{RGB}{139,69,19}   %
\definecolor{color11}{RGB}{128,0,128}   %
\definecolor{color12}{RGB}{105,105,105} %

\begin{document}

\title{\LARGE \bf Decentralized Scalar Field Mapping using Gaussian Process}
\author{
Hossein Papi$^{1}$,  Muzaffar Qureshi$^{2}$, Kyle Volle$^{3}$, 
Omkar Sudhir Patil$^{1}$, Rushikesh Kamalapurkar$^{2}$
\thanks{This research was supported in part by the Air Force Research Laboratory under contract number FA8651-24-1-0019. Any opinions, findings, or recommendations in this article are those of the author(s), and do not necessarily reflect the views of the sponsoring agencies.}
\thanks{$^{1}$ Department of Electrical and Computer Engineering, Louisiana State University, email:{\tt \footnotesize \{hpapi1, opatil1\}@lsu.edu.}}
\thanks{$^{2}$ Department of Mechanical and Aerospace Engineering, University of Florida, email: {\tt\footnotesize \{muzaffar.qureshi, rkamalapurkar\} @ufl.edu}.}
\thanks{$^{3}$ Torch Technologies, Shalimar, Florida, USA, email: {
\tt \footnotesize Kyle.Volle@torchtechnologies.com}.}
\thanks{An AI-based assistant was used for simulation,
proofreading, and language editing.  The problem formulation, mathematical
development, and rule design were carried out by the authors, who reviewed and
verified all output and take full responsibility for the manuscript.}
}
\maketitle
\thispagestyle{empty}
\pagestyle{empty}

\begin{abstract}
Decentralized Gaussian process (GP) methods provide a scalable framework for multi-agent scalar-field estimation by allowing each agent to maintain a local GP over its assigned subdomain. Because neighboring agents condition their GP models on different measurements, their GP posteriors generally disagree over overlapping regions. This paper introduces the inducing points decentralized Gaussian process (IPD-GP) framework, which exploits these discrepancies to improve posterior consistency over shared regions. Each sender constructs receiver-specific data packets by selecting inducing locations targeted at the pairwise overlap and transmitting the corresponding posterior means and variances. Each receiver then selectively assimilates informative packets as persistent fictitious measurements in its local GP. The developed architecture preserves decentralized computation and communication without requiring full dataset exchange or centralized inference. Performance is evaluated using normalized root-mean-square error, negative log predictive density, and their overlap-restricted counterparts, with comparisons against non-communicating local GPs, decentralized aggregation methods, and a centralized GP. Simulation results show that IPD-GP achieves competitive predictive accuracy while reducing inter-agent communication.
\end{abstract}

\section{Introduction}
Mapping unknown scalar fields is fundamental to many autonomy applications, including surveillance, wildfire monitoring, and plume monitoring \cite{SCC.Sabella.1998,SCC.Lin.Liu.ea2019,SCC.Ren.Li.ea2021}. Gaussian processes (GPs) are well suited to such applications because they provide flexible nonparametric regression together with principled uncertainty quantification \cite{SCC.Rasmussen.Williams2006,SCC.Qureshi.Ogri.ea2024,Lederer2019Uniform}. They have consequently been used extensively for multi-robot environmental sampling, adaptive coverage, and active sensing \cite{Ma2018Sampling,Suryan2020MinTime,Santos2021Coverage,Jang2020ActiveSensing,Brunke2022SafeLearning}. However, GP inference becomes increasingly expensive as the number of measurements grows. For a dataset of size $\mathrm{N}$, exact GP training requires $\mathcal{O}(\mathrm{N}^3)$ computation and $\mathcal{O}(\mathrm{N}^2)$ storage, which can become prohibitive for mobile sensing teams operating with limited onboard computation and communication resources \cite{Liu2020Survey,Gielis2022Critical,SCC.Ogri.Qureshi.ea2025a}.

Decentralized scalar-field estimation addresses these limitations by distributing sensing and inference among multiple agents. In this setting, each agent collects measurements within a local subdomain and maintains its own GP posterior \cite{qureshi2025gaussianprocessbasedscalarfield}. When neighboring subdomains overlap, the corresponding agents generally obtain different posterior means and variances over the shared region because their GPs are conditioned on different datasets. Existing distributed estimation methods reduce such discrepancies through consensus, distributed fusion, or continuity constraints \cite{Cortes2009Kriged,Park2018Patchwork}. In contrast, this paper treats posterior disagreement over an overlap as an indication that neighboring agents possess complementary information. This observation motivates selective exchange of posterior information specifically over shared regions, rather than reconstruction of a centralized posterior or global consensus among all local models.

Several classes of scalable GP methods are related to this objective. Aggregation-of-experts methods partition data among local GP models and combine their predictive distributions at query locations. Representative approaches include products of experts (PoE) \cite{Hinton2002PoE,Cao2014gPoE}, Bayesian committee machines (BCM) \cite{Tresp2000BCM,Deisenroth2015Distributed,liu2018generalizedrobustbayesiancommittee}, and nested Kriging \cite{Rulliere2018Nested}. Decentralized implementations replace a central fusion mechanism with consensus or distributed computation \cite{Cortes2009Kriged,Chen2015Fusion,Pillonetto2019Finite,Kontoudis2021Nested,Yuan2024Lightweight}, and some approaches further restrict participating agents using covariance-based criteria \cite{muzaffar.Kontoudis.Stilwell2026}. These methods are primarily query-centric: they determine how local predictions or participating agents are combined at a requested test location. In contrast, the problem considered here is communication-centric and geometry-aware: a sender determines which portion of its posterior information should be communicated to a particular receiver based on their pairwise overlap.

Other approaches address complementary aspects of scalability. Federated GP methods distribute hyperparameter learning or model training across agents \cite{Kontoudis2024Federated,Xie2019ADMM,Yue2024FederatedGP}, whereas the present work focuses on prediction-time information exchange between already established local GP models. Sparse GP methods reduce computational cost using pseudo-inputs or inducing points \cite{QuinoneroCandela2005Unifying,Snelson2005Pseudo,Titsias2009Variational,Datta2016NNGP,Gramacy2015LocalGP,Cole2021LIGP,Hoang2019Collective,Lederer2021RealTime}. These methods generally select compact representations to approximate a GP model or improve prediction over prescribed targets. Here, inducing points instead serve as communication objects whose locations are selected specifically for each sender-receiver overlap. Patchwork kriging \cite{Park2018Patchwork} is closely related in its use of pseudo-observations to improve consistency between neighboring local models. However, it imposes continuity through a predefined boundary construction, whereas the proposed method generates posterior-derived information online and selectively assimilates it according to the current local models.

Motivated by this gap, this paper develops the inducing-points decentralized Gaussian process (IPD-GP) framework for communication-efficient scalar-field mapping with overlapping local GP models. The main contributions are:

\begin{itemize}
    \item An overlap-aware decentralized information-sharing framework tailored to pairwise overlap regions.

    \item A receiver-specific inducing-point communication strategy with local packet assimilation.

    \item Persistent assimilation of accepted neighbor information into subsequent local GP predictions.

    \item Evaluation against local, decentralized aggregation, and centralized GP baselines using prediction accuracy and uncertainty.
\end{itemize}

\section{Problem Formulation}\label{sec:ProblemFormulation}

Table \ref{tab:notation} summarizes the key notation used throughout this paper. 

Consider a collection of $N$ mobile sensing agents, indexed by $\mathcal{V}=\{1,\dots,N\}$, which communicate over a connected directed graph $\mathcal{G}=(\mathcal{V},\mathcal{E})$, where $\mathcal{E}\subseteq \mathcal{V}\times\mathcal{V}$ denotes the set of communication links. For each agent $i\in\mathcal{V}$, define the out-neighbor set as $\mathcal{N}_{i}^{\mathrm{out}}\coloneqq \{\, j\in\mathcal{V}\mid (i,j)\in\mathcal{E}\,\},$
and the in-neighbor set as $\mathcal{N}_{i}^{\mathrm{in}}\coloneqq \{\, j\in\mathcal{V}\mid (j,i)\in\mathcal{E}\,\}.$

Given a compact domain $\mathcal{X} \subset \mathbb{R}^n$, let $f: \mathcal{X} \rightarrow \mathbb{R}$ denote an unknown scalar field representing the physical quantity of interest.  To distribute the mapping task across the collection of agents, each agent $i \in \mathcal{V}$ is assigned a subdomain $\mathcal{X}^i \subseteq \mathcal{X}$ for measurements, and these subdomains collectively cover the domain, namely, $\mathcal{X} \subseteq \bigcup_{i\in\mathcal{V}} \mathcal{X}^i.$
For any pair of agents $i,j \in \mathcal{V}$, define the overlap region $\mathcal{O}^{ij} \coloneqq \mathcal{X}^i \cap \mathcal{X}^j.$

The agents evolve on a common synchronized measurement clock indexed by $t\in\mathbb{N}$. Let $\mathcal{D}_{t}^{i}\subseteq \mathcal{X}^{i}\times\mathbb{R}$ denote the set of measurements collected by agent $i$ up to clock step $t$. Each local measurement pair $(x^{i},y^{i})\in \mathcal{D}_{t}^{i}$ satisfies 
$y^{i}=f\!\left(x^{i}\right)+\epsilon^{i},$
where $x_{\tau}^{i}\in\mathcal{X}^{i}$ and $\epsilon^{i}\sim\mathcal{N}\!\left(0,(\sigma^{i})^{2}\right)$. The noise variables are assumed to be independent across agents and independent and identically distributed (i.i.d.) over $\tau$ for each fixed agent $i$. In addition to its local measurements \(\mathcal{D}_{t}^{i}\) of cardinality \(S_{t}^{i}\), agent \(i\) maintains a set of fictitious measurements \(\mathcal{M}_{t}^{i}\) representing packets selected for assimilation from its neighbors at prior time steps. Let the augmented information set of agent $i$ be defined as $\mathcal{A}_{t}^{i} \coloneqq \left( \mathcal{D}_{t}^{i}, \mathcal{M}_{t}^{i} \right)$, and let $\mathcal{A}$ denote the space of all such possible augmented information sets. The local GP of each agent is written as 
\begin{equation}
f \mid \mathcal{A}_{t}^{i} \sim \mathcal{GP}\!\left( \mu_{t}^{i}(\cdot), \Sigma_{t}^{i}(\cdot,\cdot) \right),
\end{equation}
where \(\mu_{t}^{i}:\mathcal{X}\to\mathbb{R}\) is the posterior mean function and \(\Sigma_{t}^{i}:\mathcal{X}\times\mathcal{X}\to\mathbb{R}\) is the posterior covariance function. Based on this local GP, agents exchange information. Let the space of all valid communication packets be defined as $\mathfrak{U} \coloneqq \mathcal{X} \times \mathbb{R} \times \mathbb{R}_{\geq 0}$. The transmission rule used by agent $j$ to generate data for agent $i$ is formally defined as a mapping $\Phi_{t}^{j\to i} : \mathcal{A} \times \mathcal{P}(\mathcal{X}) \to \mathcal{P}(\mathfrak{U})$, where $\mathcal{P}(\cdot)$ denotes the power set. Each in-neighbor $j$ evaluates its local augmented data $\mathcal{A}_t^{j}$ and the overlap geometry $\mathcal{O}^{ji}$ to construct a candidate packet library $\mathcal{C}_{t}^{j\to i} = \Phi_{t}^{j\to i}\!\left(\mathcal{A}_{t}^{j},\mathcal{O}^{ji}\right) \subset \mathfrak{U}$ to transmit to agent $i$.
Each generic datum in $\mathcal{C}_{t}^{j\to i}$, referred to hereafter as a packet, has the form
\begin{equation}
\mathbf{u} = (u, m, s) \in \mathfrak{U} \coloneqq \mathcal{X} \times \mathbb{R} \times \mathbb{R}_{\geq 0},
\label{eq:packet_definition}
\end{equation}where \(u\) is a spatial location, \(m\) is the predictive posterior mean of the unknown scalar field $f$ evaluated at $u$ by the sender's local GP, and $s$ is the corresponding predictive posterior variance at $u$. By using a generic packet notation $\mathbf{u}$, we distinguish candidate packets from the indexed packets retained in $\mathcal{M}_{t}^{i}$. The data received by agent \(i\) from all of its in-neighbors is pooled into the candidate packet library, defined as $\bar{\mathcal{C}}_{t}^{i} \coloneqq \bigcup_{j\in\mathcal{N}_{i}^{\mathrm{in}}} \mathcal{C}_{t}^{j\to i}.$

Agent \(i\) applies a local selection rule 
$\psi_{t}^{i}:\mathcal{P}(\mathfrak{U})\to \mathfrak{U},$
which maps the pooled candidate library to one selected packet to be assimilated into its local GP model. Let
\begin{equation}
\mathcal{M}_{t}^{i} \coloneqq\left\{\left(u_{\ell,t}^{i},m_{\ell,t}^{i},r_{\ell,t}^{i}\right) \right\}_{\ell=1}^{L_t^i} \subseteq \mathcal{X}\times\mathbb{R}\times\mathbb{R}_{>0}
\label{eq:fictitious_measurements_i}
\end{equation}
denote the set of packets selected for assimilation by agent \(i\) up to clock step \(t\). The local GP maintained by agent \(i\) is then conditioned on both its local measurements \(\mathcal{D}_{t}^{i}\) of cardinality \(S_{t}^{i}\) and the packets \(\mathcal{M}_{t}^{i}\) of cardinality \(L_{t}^{i}\) selected for assimilation from the data received from their neighbors.

At each clock step $t$, every agent acquires one new local measurement, yielding a local measurement budget $S_t^i = t$. The transmission budget on each edge $(j,i)$ is constrained to one packet comprising $p_t^{j \to i}$ inducing points and their moments, requiring the sender to transmit $p_t^{j \to i}$ scalars per step. To prevent unbounded growth in computation, the receiver-side assimilation budget is restricted to one fictitious measurement per clock step, meaning $L_t^i \le t$. Consequently, while each agent's local GP training and prediction costs scale as $\mathcal{O}((S_t^i+L_t^i)^3)$ and $\mathcal{O}((S_t^i+L_t^i)^2)$ respectively, the inter-agent communication cost remains constant and is significantly lower than transmitting full local datasets. IPD-GP performance is measured by overlap root-mean-square error (RMSE), negative log predictive density (NLPD), and accuracy-to-communication cost. 

The objective of this paper is to test the hypothesis that the discrepancy between the local GPs regarding the overlaps of their sub-domains provides a useful decentralized metric that can be used to improve the predictive accuracy of the entire domain of the group. To that end, we design novel decentralized metrics, based on the consistency of the local GPs on subdomain overlaps, that guide the design of the transmission rules $\Phi_t^{j\to i}$ and the selection rules $\psi_t^{i}$. We first describe the local GP models associated with the local measurements \(\mathcal{D}_t^i\), then introduce a sparse inducing-point approximation to generate reduced-order representations of the local GPs, and finally develop an inducing-point selection rule that determines which communication packets should be transmitted and assimilated by the agents. 

\begin{remark}
The subdomains $\mathcal{X}^i$, and hence the overlap geometry
$\mathcal{O}^{ij}$, are fixed for all $t$; agent motion affects only where
measurements are taken within $\mathcal{X}^i$.
\end{remark}

{\begin{table}[t]\centering\footnotesize
\caption{Notation.}\label{tab:notation}
\begin{tabular}{@{}ll@{}}\hline
$\mathcal{X}^i,\ \mathcal{O}^{ij}$ & subdomain of agent $i$; overlap $\mathcal{X}^i\cap\mathcal{X}^j$\\
$\mathcal{D}_t^i,\ S_t^i$ & local measurement set; its cardinality\\
$\mathcal{M}_t^i,\ L_t^i$ & assimilated packet set; its cardinality\\
$\mathcal{A}_t^i$ & augmented information set $(\mathcal{D}_t^i,\mathcal{M}_t^i)$\\
$R_t^i$ & diagonal noise matrix of the augmented set\\
$\alpha^i,\ \ell^i$ & signal scale and length-scale of agent $i$\\
$\nu^j,\ \theta^j$ & $(\alpha^j)^2$ and $2(\ell^j)^2$\\
$\mathcal{T}_t^{j\to i},\ q_t^{j\to i}$ & target set in $\mathcal{O}^{ji}$; its cardinality\\
$\alpha_{b,t}^{j\to i}$ & weight of target $z_{b,t}^{j\to i}$, $\sum_b\alpha_{b,t}^{j\to i}=1$\\
$p_t^{j\to i}$ & number of inducing points sent on edge $j\to i$\\
$\lambda_\mu,\ \lambda_\Sigma$ & mean / variance weights in $J_t^i$\\
$a_\ell^{j\to i},\ b_\ell^{j\to i}$ & bounds of the box enclosing $\mathcal{O}^{ji}$\\ \hline
\end{tabular}
\vspace{-0.6cm}
\end{table}

\section{Methods}
This section illustrates the local GP models maintained by each agent from its augmented information set. Later Section~\ref{sec:batch_targeted_ip_selection} describes the optimal selection of inducing points to construct targeted communication packets for neighboring agents.
\subsection{Gaussian Process Regression}
\label{sec:GP_section}
We first make precise the local GP model that each agent \(i\in\mathcal{V}\) generates from its augmented information set \(\mathcal{A}_t^i=(\mathcal{D}_t^i,\mathcal{M}_t^i)\). This GP model is the object that the transmission rule \(\Phi_t^{j\to i}\) and the selection rule \(\psi_t^i\) are ultimately intended to improve. The raw dataset \(\mathcal{D}_t^i\) contains locally collected measurements, while \(\mathcal{M}_t^i\) contains packets selected for assimilation from received packet libraries $\bar{\mathcal{C}}_{t}^{i}$. Each agent \(i\) employs a squared exponential covariance kernel of the form 
\begin{equation}
k^{i}(x,x') = (\alpha^{i})^{2} \exp\!\left( -\frac{\|x-x'\|^{2}}{2(\ell^{i})^{2}} \right), \qquad x,x'\in\mathcal{X},\end{equation} where \(\alpha^{i}>0\) and \(\ell^{i}>0\) denote the signal variance and length-scale hyperparameters of agent \(i\), respectively.

Let $S_{t}^{i}\coloneqq \left|\mathcal{D}_{t}^{i}\right|$ and $L_{t}^{i}\coloneqq
\left|\mathcal{M}_{t}^{i}\right|$ denote the number of local measurements and
the number of assimilated packets, respectively, and write the local
measurements of agent \(i\) at clock step \(t\) as $\mathcal{D}_{t}^{i} =
\left\{\left(x_{j,t}^{i},y_{j,t}^{i}\right)\right\}_{j=1}^{S_{t}^{i}}$, where
\(x_{j,t}^{i}\in\mathcal{X}^{i}\) and \(y_{j,t}^{i}\in\mathbb{R}\). Likewise,
write the packets selected for assimilation as
\eqref{eq:fictitious_measurements_i} where \(u_{\ell,t}^{i}\in\mathcal{X}\),
\(m_{\ell,t}^{i}\in\mathbb{R}\), and \(r_{\ell,t}^{i}>0\) are the location,
mean and variance estimates associated with the selected packet, respectively. For compactness of notation, define
$X_{\mathrm{raw},t}^{i}\coloneqq \left[ x_{1,t}^{i},\dots,x_{S_t^i,t}^{i} \right]^{\top},\ Y_{\mathrm{raw},t}^{i} \coloneqq \left[ y_{1,t}^{i},\dots,y_{S_t^i,t}^{i} \right]^{\top}, \ U_{t}^{i}\coloneqq \left[ u_{1,t}^{i},\dots,u_{L_t^i,t}^{i} \right]^{\top}, \ M_{t}^{i,\mathrm{fic}} \coloneqq \left[ m_{1,t}^{i},\dots,m_{L_t^i,t}^{i} \right]^{\top},$ and the augmented input and output arrays $X_{t}^{i} \coloneqq \left[ X_{\mathrm{raw},t}^{i},  U_t^i \right]^\top, \ Y_{t}^{i} \coloneqq \left[ Y_{\mathrm{raw},t}^{i}, \ M_{t}^{i,\mathrm{fic}} \right]^\top$. Finally, define the associated diagonal noise matrix
\begin{equation}
R_t^i
\coloneqq
\operatorname{diag}
\left(
(\sigma^i)^2,\dots,(\sigma^i)^2, r_{1,t}^{i},\dots,r_{L_t^i,t}^{i}
\right).
\label{eq:augmented_noise_matrix}
\end{equation}

To facilitate decentralized assimilation, every element of the augmented set
$\mathcal{A}_t^i$ is treated as a conditionally independent observation of the
latent field, which is what renders $R_t^i$ in
\eqref{eq:augmented_noise_matrix} diagonal. Each transmitted value $m_{\ell,t}^{i}$ is a sender posterior mean rather than a
direct sample of $f$, so its error is a functional of the sender's dataset and
packets originating from a common sender remain mutually correlated even after
conditioning on the field.  Representing that dependence would require
the cross-covariance between datasets of two agents, reinstating the raw-data
exchange the packet interface is designed to avoid, and the same device
underlies the diagonal correction $\Lambda_{t}^{i,(p)}$ in
\eqref{eq:Sigma_inducing_i}
\cite{QuinoneroCandela2005Unifying,Snelson2005Pseudo}.  The cost is that
$\left(\sigma_t^i(x^*)\right)^2$ under-states the true posterior variance when
admitted packets are correlated, and the budget of one packet per clock step
bounds the rate at which that error accumulates.

Given the augmented information set \(\mathcal{A}_{t}^{i}\), define the kernel vector \(\mathbf{k}_{t}^{i}(x^{*})\in\mathbb{R}^{S_t^i+L_t^i}\) and kernel matrix \(K_{t}^{i}\in\mathbb{R}^{(S_t^i+L_t^i)\times(S_t^i+L_t^i)}\) by 
$\mathbf{k}_{t}^{i}(x^{*}) \coloneqq \left[ k^{i}(x^{*},X_t^i(1,:)), \dots, k^{i}(x^{*},X_t^i(S_t^i+L_t^i,:)) \right]^{\top},$
$\left[K_{t}^{i}\right]_{j,\ell} \coloneqq k^{i}(X_t^i(j,:),X_t^i(\ell,:)), \ j,\ell=1,\dots,S_t^i+L_t^i.$

Using \(\mathbf{k}_{t}^{i}(x^{*})\), \(K_{t}^{i}\), and the noise matrix \(R_t^i\) in \eqref{eq:augmented_noise_matrix}, the posterior mean and variance of agent \(i\) are computed as
\begin{align}
\mu_{t}^{i}(x^{*}) = {\mathbf{k}_{t}^{i}(x^{*})}^{\top} \left( K_{t}^{i}+R_t^i \right)^{-1} Y_{t}^{i},
 \text{ and} \\
\scalebox{0.9}{$
\left(\sigma_{t}^{i}(x^{*})\right)^{2} = k^{i}(x^{*},x^{*}) - {\mathbf{k}_{t}^{i}(x^{*})}^{\top} \left( K_{t}^{i}+R_t^i \right)^{-1} \mathbf{k}_{t}^{i}(x^{*}),$}
\end{align}
where the first \(S_t^i\) diagonal entries of \(R_t^i\) correspond to local sensor-noise variance, and the remaining \(L_t^i\) diagonal entries correspond to the retained pseudo-measurement variances associated with previously selected packets.

Note that the computational cost of the GP prediction grows cubically with the total size of the augmented information set. In particular, since agent \(i\) conditions on \(S_t^i\) raw measurements and \(L_t^i\) retained fictitious measurements, each evaluation of the posterior mean requires \(\mathcal{O}\!\left((S_t^i+L_t^i)^3\right)\) time, and each evaluation of the posterior variance requires \(\mathcal{O}\!\left((S_t^i+L_t^i)^2\right)\) time. To reduce this burden while preserving the agent-specific local GP structure, the next subsection introduces an inducing-point approximation for each agent.

\subsection{Inducing Points in Gaussian Process}
\label{sec:inducing_points}
To obtain a scalable local approximation of a local GP, fix an arbitrary agent \(i\in\mathcal{V}\) and clock step \(t\in\mathbb{N}\). Recall from Section~\ref{sec:GP_section} that the local posterior of agent \(i\) is conditioned on the augmented information set \(\mathcal{A}_t^i=(\mathcal{D}_t^i,\mathcal{M}_t^i)\), whose total size is \(S_t^i+L_t^i\). To generate a reduced order representation of its GP, agent \(i\) employs a set of \(p_t^i\) inducing points, with \(p_t^i \ll S_t^i+L_t^i\) \cite{SCC.Rasmussen.Williams2006}.

Specifically, let 
\begin{equation}\scalebox{0.96}{$
\bar{X}_{t}^{i} \coloneqq \left[ \bar{x}_{1,t}^{i}, \dots, \bar{x}_{p_t^i,t}^{i} \right]^\top \subset \mathcal{X}^{i},$}
\end{equation} denote the inducing-point set of agent \(i\) at clock step \(t\). Using the kernel \(k^i(\cdot,\cdot)\), define 
$\left[ K_{pp,t}^{i} \right]_{r,s} \coloneqq k^{i}(\bar{x}_{r,t}^{i}, \bar{x}_{s,t}^{i}), \ r,s = 1,\dots,p_t^i,$ and $\left[ K_{Xp,t}^{i} \right]_{j,r} \coloneqq k^{i}\!\left( X_t^i(j,:), \bar{x}_{r,t}^{i} \right), j = 1,\dots,S_t^i+L_t^i, \ r = 1,\dots,p_t^i,$ where \(X_t^i(j,:)\) denotes the \(j\)-th input location in the augmented input array \(X_t^i\), while \(\bar{x}_{r,t}^{i}\) denotes the \(r\)-th inducing point selected for the sparse approximation.

The corresponding low-rank approximation of the training covariance is
\begin{equation}
K_{t}^{i} \approx \tilde{K}_{t}^{i} \coloneqq K_{Xp,t}^{i} \left( K_{pp,t}^{i} \right)^{-1} \left( K_{Xp,t}^{i} \right)^\top.
\end{equation}
To equate the exact variance, we use the diagonal correction
\begin{equation}
\Sigma_{t}^{i,(p)}
\coloneqq
K_{Xp,t}^{i}
\left( K_{pp,t}^{i} \right)^{-1}
\left( K_{Xp,t}^{i} \right)^\top
+
\Lambda_{t}^{i,(p)}
+
R_t^i,
\label{eq:Sigma_inducing_i}
\end{equation}
where $\Lambda_{t}^{i,(p)} \coloneqq \operatorname{Diag} \left\{ K_{t}^{i} - K_{Xp,t}^{i} \left( K_{pp,t}^{i} \right)^{-1} \left( K_{Xp,t}^{i} \right)^\top \right\}.$ For a test point \(x^* \in \mathcal{X}^{i}\), define 
$k_{p,t}^{i}(x^{*})
\coloneqq
\left[
k^{i}(\bar{x}_{1,t}^{i},x^{*}),
\dots,
k^{i}(\bar{x}_{p_t^i,t}^{i},x^{*})
\right]^\top,$ 
$\Omega_{t}^{i}
\coloneqq
\Lambda_{t}^{i,(p)} + R_t^i,$ 
$\Gamma_{t}^{i}
\coloneqq
\left(\Omega_{t}^{i}\right)^{-1} K_{Xp,t}^{i},$
and $Q_{t}^{i}
\coloneqq
K_{pp,t}^{i}
+
\left(K_{Xp,t}^{i}\right)^\top
\left(\Omega_{t}^{i}\right)^{-1}
K_{Xp,t}^{i}.$
The resulting sparse GP predictive mean and variance are
\begin{equation}
\mu_{t}^{i,(p)}(x^{*})
=
\left(k_{p,t}^{i}(x^{*})\right)^\top
\left(Q_{t}^{i}\right)^{-1}
\left(\Gamma_{t}^{i}\right)^\top
Y_{t}^{i},
\label{eq:mu_inducing_i}
\end{equation}
\begin{multline}
\left(\sigma_{t}^{i,(p)}(x^{*})\right)^{2}
=
k^{i}(x^{*},x^{*})
\\
-
\left(k_{p,t}^{i}(x^{*})\right)^{\top}
\left[
\left(K_{pp,t}^{i}\right)^{-1}
-
\left(Q_{t}^{i}\right)^{-1}
\right]
k_{p,t}^{i}(x^{*}).
\label{eq:sigma_inducing_i}
\end{multline}
The inducing-point construction reduces the dominant local training cost from \(\mathcal{O}\!\left((S_t^i+L_t^i)^3\right)\) to \(\mathcal{O}\!\left((S_t^i+L_t^i)(p_t^i)^2 + (p_t^i)^3\right)\), and the storage cost from \(\mathcal{O}\!\left((S_t^i+L_t^i)^2\right)\) to \(\mathcal{O}\!\left((S_t^i+L_t^i)p_t^i + (p_t^i)^2\right)\label{eq:cost}.\)

The computational calculation in decentralized GP inference has two components: the local cost incurred by each agent for training, storage, and prediction as its local augmented dataset evolves, and the communication cost required to transmit candidate packets between agents. Because received information increases the computational load of the receiving agent and incurs a high communication cost, direct exchange of exact GP representations is not scalable. 

In this paper, we adopt the idea of inducing points to facilitate the design of the transmission rule $\Phi_{t}^{j\to i}$. Commensurate with the central hypothesis of the paper, the transmission rules are designed to reduce the mismatch between the local GPs on the overlapping subdomains. To that end, each sender selects a small set of inducing points for a specific receiver so that the transmitted packets are the most informative over their overlap region $\mathcal{O}^{ji}$. In the next subsection, an edge-dependent inducing-point selection rule is developed for communication-efficient inter-agent overlap consistency.

\section{Domain Targeted Inducing-Point Selection}
\label{sec:batch_targeted_ip_selection}
In Section~\ref{sec:ProblemFormulation}, the transmission rule \(\Phi_t^{j\to i}\) was introduced abstractly as a map from agent \(j\)'s retained information and the overlap geometry to a candidate set sent to agent \(i\). Here, we realize that idea by constructing edge-specific inducing points that are selected not to summarize the entirety of agent \(j\)'s subdomain, but to reduce predictive uncertainty over the portion of \(\mathcal{O}^{ji}\) that is most relevant to agent \(i\). Fix a communicating pair \((j,i)\in\mathcal{E}\) and a clock step \(t\in\mathbb{N}\). Let \(N_t^j \coloneqq S_t^j+L_t^j\) denote the total size of the augmented information set of sender agent \(j\). Let \(p_t^{j\to i}\in\mathbb{N}\) denote the total number of edge-specific inducing points to be selected by sender agent \(j\) for receiver agent \(i\) at clock step \(t\), and let $m\in\{0,1,\dots,p_t^{j\to i}-1\}$ denote the current greedy selection stage. The localized inducing-point method of \cite{Cole2021LIGP} selects inducing points for a single predictive location by minimizing a weighted integrated prediction-error criterion. Here, that construction is extended from a single target location to a finite target set, so that agent \(j\) selects edge-specific inducing points to improve prediction collectively over the portion of the overlap region that is most relevant to agent \(i\).

For notational convenience, write the squared exponential kernel of agent \(j\) as $k^{j}(x,x') = \nu^{j}\kappa^{j}(x,x'), \ \nu^{j}\coloneqq(\alpha^{j})^{2}, \ \kappa^{j}(x,x') \coloneqq \exp\!\left( -\frac{\|x-x'\|^{2}}{\theta^{j}} \right),
\ \theta^{j}\coloneqq 2(\ell^{j})^{2},$ where \(\alpha^{j}\) and \(\ell^{j}\) are the kernel hyperparameters already introduced in Section~\ref{sec:GP_section}.

Let $\mathcal{T}_{t}^{j\to i} = \left\{
z_{1,t}^{j\to i},\dots,z_{q_t^{j\to i},t}^{j\to i} \right\} \subseteq \mathcal{O}^{ji}$ denote the evaluation set in the overlap region, where \(q_t^{j\to i}\in\mathbb{N}\) is the number of target locations used for communication from agent \(j\) to agent \(i\) at clock step \(t\). Let $\bar{X}_{m,t}^{j\to i}
=
\left[
\bar{x}_{1,t}^{j\to i},\dots,\bar{x}_{m,t}^{j\to i}
\right]^{\top}
\subseteq \mathcal{O}^{ji}$ be the current ordered edge-specific inducing-point set after \(m\) selections, and let $\bar{x}_{m+1,t}^{j\to i}
=
\left[
\bar{x}_{m+1,1,t}^{j\to i},\dots,\bar{x}_{m+1,n,t}^{j\to i}
\right]^{\top}
\in \mathcal{O}^{ji}$ denote a candidate next inducing point. The corresponding augmented inducing-point set is $\bar{X}_{m+1,t}^{j\to i}
=
\left[
\bar{x}_{1,t}^{j\to i},\dots,\bar{x}_{m,t}^{j\to i},\bar{x}_{m+1,t}^{j\to i}
\right]^{\top}.$

Let $\tilde{x}\in\mathcal{O}^{ji}$ denote a generic evaluation location in the
overlap region.  It is the variable of integration in the criterion developed
below and is integrated out in \eqref{eq:BTIP_integral}, so it is neither a
test point nor a design variable.  To evaluate the utility of
\(\bar{x}_{m+1,t}^{j\to i}\) over the full target set \(\mathcal{T}_{t}^{j\to i}\),
define the weighting function
\begin{equation}\scalebox{0.98}{$
\omega_{\mathcal{T}_{t}^{j\to i}}(\tilde{x})
=
\sum_{b=1}^{q_t^{j\to i}}
\alpha_{b,t}^{j\to i}\,
\kappa^{j}\!\left(\tilde{x},z_{b,t}^{j\to i}\right),$}
\label{eq:domain_weight_function}
\end{equation}
where \(\alpha_{b,t}^{j\to i}\ge 0\) specifies the relative importance of the target location \(z_{b,t}^{j\to i} \), such that $\sum_{b=1}^{q_t^{j\to i}}\alpha_{b,t}^{j\to i}=1$. The target set enters the criterion only through
$\omega_{\mathcal{T}_{t}^{j\to i}}$, which is where it is meant to act.  The
integral sweeps the whole overlap through $\tilde{x}$ and the weight
concentrates that measure near the targets, so $\mathcal{T}_{t}^{j\to i}$
specifies where in $\mathcal{O}^{ji}$ predictive accuracy matters to the
receiver rather than where the criterion is evaluated.  In the simulations it
is the set of grid nodes in $\mathcal{O}^{ji}$ with uniform weights
$\alpha_{b,t}^{j\to i}=1/q_{t}^{j\to i}$.

Next, define the kernel vector $k_{m+1,t}^{j\to i}(x)
\coloneqq
\left[
k^{j}(x,\bar{x}_{1,t}^{j\to i}),
\dots,
k^{j}(x,\bar{x}_{m+1,t}^{j\to i})
\right]^{\top} \in\mathbb{R}^{m+1},$ the Gram matrix $\left[K_{m+1,t}^{j\to i}\right]_{r,s}
\coloneqq
k^{j}(\bar{x}_{r,t}^{j\to i},\bar{x}_{s,t}^{j\to i}), \ r,s=1,\dots,m+1,$ and the cross-covariance matrix between the sender active inputs \(X_t^j=\{x_{u,t}^{j}\}_{u=1}^{N_t^j}\) and the augmented edge-specific inducing set $\left[K_{X,m+1,t}^{j\to i}\right]_{u,r}
\coloneqq
k^{j}(X_t^j(u,:),\bar{x}_{r,t}^{j\to i}), \ u=1,\dots,N_t^j.$

Using the augmented training covariance matrix \(K_t^j\) from Section~\ref{sec:GP_section}, define $\scalebox{0.9}{$
\Lambda_{m+1,t}^{j\to i}
\coloneqq
\operatorname{Diag}
\left\{
K_t^j
-
K_{X,m+1,t}^{j\to i}
\left(K_{m+1,t}^{j\to i}\right)^{-1}
\left(K_{X,m+1,t}^{j\to i}\right)^{\top}
\right\},$}$ $\Omega_{m+1,t}^{j\to i}
\coloneqq
\Lambda_{m+1,t}^{j\to i}
+
R_t^j,$ and $Q_{m+1,t}^{j\to i}
\coloneqq
K_{m+1,t}^{j\to i}
+
\left(K_{X,m+1,t}^{j\to i}\right)^{\top}
\left(\Omega_{m+1,t}^{j\to i}\right)^{-1}
K_{X,m+1,t}^{j\to i}.$

The sparse predictive variance of sender agent \(j\), computed from the augmented information set \(\mathcal{A}_t^j\) and the augmented edge-specific inducing set \(\bar{X}_{m+1,t}^{j\to i}\), evaluated at any $\tilde{x}\in\mathcal{O}^{ji}$, is
\begin{multline}
\left(\sigma_{m+1,t}^{j\to i}(\tilde{x})\right)^{2}
=
k^{j}(\tilde{x},\tilde{x})
-
\left(k_{m+1,t}^{j\to i}(\tilde{x})\right)^{\top} \\
\left[
\left(K_{m+1,t}^{j\to i}\right)^{-1}
-
\left(Q_{m+1,t}^{j\to i}\right)^{-1}
\right]
k_{m+1,t}^{j\to i}(\tilde{x}).\end{multline}

We now define the batch-targeted inducing-point criterion
\begin{multline}
\operatorname{BTIP}_{t}^{j\to i,(m+1)}
\!\left(
\bar{x}_{m+1,t}^{j\to i};
\mathcal{T}_{t}^{j\to i}
\right)
\\
=
\int_{\mathcal{O}^{ji}}
\omega_{\mathcal{T}_{t}^{j\to i}}(\tilde{x})\,
\frac{\left(\sigma_{m+1,t}^{j\to i}(\tilde{x})\right)^{2}}{\nu^{j}}
\,d\tilde{x}.
\label{eq:BTIP_integral}
\end{multline}
Since \(k^{j}(\tilde{x},\tilde{x})=\nu^{j}\), the normalization by \(\nu^{j}\) removes the process scale from the variance term. Although \eqref{eq:BTIP_integral} is induced by the predictive variance, we interpret it as a weighted spatially integrated surrogate of prediction error over the target region. Thus, minimizing it selects inducing points that preferentially reduce posterior uncertainty in regions emphasized by \(\omega_{\mathcal{T}_{t}^{j\to i}}\). Substituting \eqref{eq:domain_weight_function} into \eqref{eq:BTIP_integral} gives
\begin{multline}
\operatorname{BTIP}_{t}^{j\to i,(m+1)}
\!\left(
\bar{x}_{m+1,t}^{j\to i};
\mathcal{T}_{t}^{j\to i}
\right)
\\
=
\sum_{b=1}^{q_t^{j\to i}}
\alpha_{b,t}^{j\to i}
\int_{\mathcal{O}^{ji}}
\kappa^{j}\!\left(\tilde{x},z_{b,t}^{j\to i}\right)
\frac{\left(\sigma_{m+1,t}^{j\to i}(\tilde{x})\right)^{2}}{\nu^{j}}
\,d\tilde{x}.
\label{eq:BTIP_sum_form}
\end{multline}

\begin{assumption}\label{as:box}
For each edge $(j,i)$, the overlap region is the axis-aligned box
$\mathcal{O}^{ji}=\prod_{\ell=1}^{n}\left[a_{\ell}^{j\to i},\,b_{\ell}^{j\to i}\right]$,
where $a_{\ell}^{j\to i}$ and $b_{\ell}^{j\to i}$ are the minimum and maximum of
the $\ell$th coordinate over $\mathcal{O}^{ji}$.
\end{assumption}

For each target point \(z \in \mathcal{O}^{ji}\), define the matrix $W_{m+1,t}^{\ast,j\to i}(z) \in  \mathbb{R}^{(m+1)\times(m+1)}$ whose \((r,s)\)-entry is
\begin{equation*}\scalebox{0.95}{$
\left[W_{m+1,t}^{\ast,j\to i}(z)\right]_{r,s}
\coloneqq
\frac{1}{\nu^{j}}
\int_{\mathcal{O}^{ji}}
\kappa^{j}(\tilde{x},z)\,
k^{j}(\tilde{x},\bar{x}_{r,t}^{j\to i})\,
k^{j}(\tilde{x},\bar{x}_{s,t}^{j\to i})
\,d\tilde{x}.$}
\end{equation*}
The corresponding batch-weighted matrix is $W_{m+1,t}^{\ast,j\to i}(\mathcal{T}_{t}^{j\to i})
=
\sum_{b=1}^{q_t^{j\to i}}
\alpha_{b,t}^{j\to i}\,
W_{m+1,t}^{\ast,j\to i}(z_{b,t}^{j\to i})$. For \(\mathcal{O}^{ji}\), the criterion in \eqref{eq:BTIP_sum_form} admits the closed-form representation 
\begin{multline}
\operatorname{BTIP}_{t}^{j\to i,(m+1)}
\!\left(
\bar{x}_{m+1,t}^{j\to i};
\mathcal{T}_{t}^{j\to i}
\right) 
=
C_{\mathcal{T}_{t}^{j\to i}}\\
-
\operatorname{tr}\!\Bigg[
\Big(
(K_{m+1,t}^{j\to i})^{-1}
-
(Q_{m+1,t}^{j\to i})^{-1}
\Big)
W_{m+1,t}^{\ast,j\to i}(\mathcal{T}_{t}^{j\to i})
\Bigg],
\label{eq:BTIP_closed_form}
\end{multline}
where
\begin{multline}
C_{\mathcal{T}_{t}^{j\to i}}
=
\sum_{b=1}^{q_t^{j\to i}}
\alpha_{b,t}^{j\to i}
\frac{\sqrt{\pi\theta^{j}}}{2}
\cdot\\
\prod_{\ell=1}^{n}
\left[
\operatorname{erf}\!\left(
\frac{z_{b,\ell,t}^{j\to i}-a_{\ell}^{j\to i}}{\sqrt{\theta^{j}}}
\right)
-
\operatorname{erf}\!\left(
\frac{z_{b,\ell,t}^{j\to i}-b_{\ell}^{j\to i}}{\sqrt{\theta^{j}}}
\right)
\right].
\label{eq:BTIP_const}
\end{multline}
where $a_{\ell}^{j\to i}$ and $b_{\ell}^{j\to i}$ are defined in Assumption~\ref{as:box}. The closed form \eqref{eq:BTIP_closed_form}-\eqref{eq:BTIP_const} is the finite-target-set
counterpart of \cite[Eq.~(10)]{Cole2021LIGP}, obtained by the same hyperrectangular
argument and derived in \cite[App.~A.1]{Cole2021LIGP}.

The next inducing point is selected greedily as $\bar{x}_{m+1,t}^{j\to i,\star}
=  
\arg\min_{\bar{x}\in\mathcal{O}^{ji}}
\operatorname{BTIP}_{t}^{j\to i,(m+1)}
\!\left(
\bar{x};
\mathcal{T}_{t}^{j\to i}
\right),$ and the edge-specific inducing-point set is updated according to $\bar{X}_{m+1,t}^{j\to i}  
= 
\left[
\bar{X}_{m,t}^{j\to i},
\bar{x}_{m+1,t}^{j\to i,\star}
\right].$ Each subproblem is nonconvex and is solved by a primal-dual interior-point
method using the analytic gradient \eqref{eq:BTIP_gradient}, with simple bounds
given by the box of Assumption~\ref{as:box}.  Stage $m=0$ is initialized at the weighted centroid of
$\mathcal{T}_{t}^{j\to i}$ and later stages at the best point of a coarse scan
over $\mathcal{T}_{t}^{j\to i}$, which keeps the search continuous rather than
restricting the inducing points to a fixed grid.  Only a local minimizer is
obtained.

For gradient-based optimization, the derivative of the objective with respect to coordinate \(\ell\in\{1,\dots,n\}\) of the candidate point \(\bar{x}_{m+1,t}^{j\to i}\) is
\begin{multline}
\frac{\partial}{\partial \bar{x}_{m+1,\ell,t}^{j\to i}}
\operatorname{BTIP}_{t}^{j\to i,(m+1)}
\!\left(
\bar{x}_{m+1,t}^{j\to i};
\mathcal{T}_{t}^{j\to i}
\right)
=
\\
-\operatorname{tr}\!\Bigg[
\Big(
\frac{\partial (K_{m+1,t}^{j\to i})^{-1}}{\partial \bar{x}_{m+1,\ell,t}^{j\to i}}
-
\frac{\partial (Q_{m+1,t}^{j\to i})^{-1}}{\partial \bar{x}_{m+1,\ell,t}^{j\to i}}
\Big)
W_{m+1,t}^{\ast,j\to i}(\mathcal{T}_{t}^{j\to i})
\Bigg]
\\
-\operatorname{tr}\!\Bigg[
\Big(
(K_{m+1,t}^{j\to i})^{-1}
-
(Q_{m+1,t}^{j\to i})^{-1}
\Big)
\frac{\partial W_{m+1,t}^{\ast,j\to i}(\mathcal{T}_{t}^{j\to i})}
{\partial \bar{x}_{m+1,\ell,t}^{j\to i}}
\Bigg],
\label{eq:BTIP_gradient}
\end{multline}
where
\begin{equation}
\frac{\partial W_{m+1,t}^{\ast,j\to i}(\mathcal{T}_{t}^{j\to i})}
{\partial \bar{x}_{m+1,\ell,t}^{j\to i}}
=
\sum_{b=1}^{q_t^{j\to i}}
\alpha_{b,t}^{j\to i}
\frac{\partial W_{m+1,t}^{\ast,j\to i}(z_{b,t}^{j\to i})}
{\partial \bar{x}_{m+1,\ell,t}^{j\to i}}.
\end{equation}
The initialization is given by the weighted centroid 
\begin{equation}
z_{c,t}^{j\to i}
=
\sum_{b=1}^{q_t^{j\to i}}
\alpha_{b,t}^{j\to i}\,
z_{b,t}^{j\to i},
\qquad
\bar{x}_{1,t}^{j\to i}=z_{c,t}^{j\to i}.
\end{equation}
After the edge-specific inducing-point set $\bar{X}_{t}^{j\to i,\star}
\coloneqq
\left[
\bar{x}_{1,t}^{j\to i,\star},\dots,\bar{x}_{p_t^{j\to i},t}^{j\to i,\star}
\right]^{\top}$ has been selected, sender agent \(j\) constructs the communicated packet library for receiver agent \(i\) by evaluating its local posterior at those locations. Specifically, the transmitted packet is 
\begin{multline*}\scalebox{0.85}{$
\mathcal{C}_{t}^{j\to i}
=
\left\{
\left(
\bar{x}_{r,t}^{j\to i,\star},
\mu_{t}^{j}\!\left(\bar{x}_{r,t}^{j\to i,\star}\right),
\Sigma_{t}^{j}\!\left(
\bar{x}_{r,t}^{j\to i,\star},
\bar{x}_{r,t}^{j\to i,\star}
\right)
\right)
\right\}_{r=1}^{p_t^{j\to i}}
\subseteq \mathfrak{U}.$}
\end{multline*}
Thus, sender agent \(j\) does not transmit the entire local dataset or its full posterior representation. Instead, it transmits only the edge-specific inducing locations selected by the batch-targeted rule together with the sender posterior mean and variance predictions at those locations. These data packets provide a novel compact communication interface tailored to the overlap region \(\mathcal{O}^{ji}\), and they are the objects from which the receiver-side candidate library is formed in the next section.

\section{Receiver-Side Candidate Inducing-Point Selection}
\label{sec:receiver_global_one_point_selection}
At every measurement step \(t\), the receiver agent \(i \in \mathcal V\) has access to its local measurements \(\mathcal D_t^i\) and the edge-specific packets \(\mathcal{C}_t^{j \to i}\subseteq \mathfrak{U}\) received from all of its in-neighbors \(j \in \mathcal N_i^{\mathrm{in}}\). Given any candidate packet $\mathbf{u} = (u ,m ,s ) \in \bar{\mathcal{C}}_t^i$ as defined in \eqref{eq:packet_definition}, receiver agent \(i\) interprets \((m,s)\) as a one-point fictitious measurement at location \(u\). The corresponding updated posterior mean is
\begin{multline}
\mu_{i,t}^{+}(x;\mathbf{u}) = 
\mu_t^i(x)
\ + 
\Sigma_t^i(x,u)
\left(
\Sigma_t^i(u,u)+s
\right)^{-1}
\\
\times
\left(
m-\mu_t^i(u)
\right),
\label{eq:updated_receiver_mean_given_packet_short}
\end{multline}
and the corresponding updated posterior covariance is
\begin{multline}
\Sigma_{i,t}^{+}(x,x';\mathbf{u})
\ = 
\Sigma_t^i(x,x')
\ - 
\Sigma_t^i(x,u)
\left(
\Sigma_t^i(u,u)+s
\right)^{-1}
\\
\times
\Sigma_t^i(u,x').
\label{eq:updated_receiver_cov_given_packet_short}
\end{multline}

The receiver agent \(i\) selects one packet from the pooled candidate library \(\bar{\mathcal{C}}_t^i\) at each $t$. For any candidate packet \(\mathbf{u} \in \bar{\mathcal{C}}_t^i\), define the local receiver cost 
\begin{multline}
J_t^i(\mathbf{u})
\coloneqq
\|\sum_{j\in\mathcal{N}_i^{\mathrm{in}}}
\sum_{\mathbf{v}=(v,m_v,s_v)\in\mathcal{C}_t^{j\to i}}\\
\Big(
\lambda_\mu
\left|
\mu_{i,t}^{+}(v;\mathbf{u})-m_v
\right|^2 
+\;
\lambda_\Sigma 
\left|
\Sigma_{i,t}^{+}(v,v;\mathbf{u})-s_v
\right|^2
\Big)\|
\label{eq:receiver_cost}
\end{multline} 
where $\lambda_\mu \ge 0$ and $\lambda_\Sigma \ge 0$ are weights for mean and variance consistency, respectively. \(J_t^i(\mathbf{u})\) measures how well the receiver posterior obtained by assimilating packet \(\mathbf{u}\) agrees, at all received packet locations, with the GP posterior of its in-neighbors. The receiver-side packet selection rule for agent $i$ is therefore given as 
\begin{equation}
\mathbf{u}_t^{i,\star} = \psi_t^i(\bar{\mathcal{C}}_{t}^{i} )=
\arg\min_{\mathbf{u}\in\bar{\mathcal{C}}_t^i}
J_t^i(\mathbf{u}).
\label{eq:local_one_point_optimization_short}
\end{equation}
After the minimizing packet 
\begin{equation}
\mathbf{u}_t^{i,\star}=(u_t^{i,\star},m_t^{i,\star},s_t^{i,\star})
\end{equation}
is selected, receiver agent \(i\) treats it as a fictitious measurement with
pseudo-noise variance $r_t^{i,\star}\coloneqq s_t^{i,\star}$. The selected packet
is admitted only if the sender is strictly more certain at $u_t^{i,\star}$ than
the receiver already is, giving the retained set
\begin{equation}
\mathcal{M}_{t+1}^{i}=
\begin{cases}
\mathcal{M}_{t}^{i}\cup\left\{\left(u_t^{i,\star},m_t^{i,\star},r_t^{i,\star}\right)\right\},
& \left(\sigma_t^i(u_t^{i,\star})\right)^{2} > s_t^{i,\star},\\[2pt]
\mathcal{M}_{t}^{i}, & \text{otherwise.}
\end{cases}
\label{eq:admission_gate}
\end{equation}
When the test fails, agent $i$ assimilates nothing at clock step $t$ and its
local posterior is unchanged.

\begin{remark}\label{rem:gate}
The two stages answer different questions: \eqref{eq:local_one_point_optimization_short}
selects the packet that best reconciles the receiver with its neighbors on the
overlap, while \eqref{eq:admission_gate} is a relative certainty test. The distinction matters because assimilation always reduces the
receiver's modeled variance at $u$, since
\eqref{eq:updated_receiver_cov_given_packet_short} gives
$\left(\sigma_{i,t}^{+}(u)\right)^{2}
=\left(\sigma_t^i(u)\right)^{2}s\big/\!\left(\left(\sigma_t^i(u)\right)^{2}+s\right)
\le\left(\sigma_t^i(u)\right)^{2}$, while displacing the mean toward $m$ with gain
$\left(\sigma_t^i(u)\right)^{2}\big/\!\left(\left(\sigma_t^i(u)\right)^{2}+s\right)$. A packet from a poorly informed sender therefore pulls the receiver's mean toward
a potentially inaccurate value and simultaneously makes it more confident about that value.
When $s_t^{i,\star}\ge\left(\sigma_t^i(u_t^{i,\star})\right)^{2}$ the sender's
estimate is no sharper than the receiver's own and therefore offers insufficient relative certainty to justify the risk under the proposed admission rule. The test compares two posterior variances at the same location
and is therefore evaluated entirely from quantities already carried in the
packet, at no additional communication cost.
\end{remark}

Because admission is tested on the minimizer of \eqref{eq:receiver_cost} rather
than imposed as a constraint on it, a step in which the most consistent packet
fails the test yields no assimilation even when other admissible candidates are
present. 
Waiting one step in such cases is a conservative heuristic.

Thus, the selected packet is not used only for a temporary posterior correction at time \(t\), but is retained as a pseudo-measurement and incorporated into the augmented information set used to form the local GP posterior at subsequent measurement steps. Collecting the selected inducing points for all agents as
\begin{equation}
\mathbf{u}_t
\ \coloneqq 
\left(
\mathbf{u}_t^1,\dots,\mathbf{u}_t^N
\right)
\ \in 
\bar{\mathcal{C}}_t^1\times\cdots\times\bar{\mathcal{C}}_t^N.
\label{eq:global_decision_variable_short}
\end{equation}
and defining the collective objective objective as
\begin{equation}
J_t(\mathbf{u}_t)
\ \coloneqq 
\sum_{i=1}^{N}
J_t^i(\mathbf{u}_t^i).
\label{eq:global_overlap_consistency_cost_short}
\end{equation}
Hence, 
\begin{equation}
\mathbf{u}_t^\star \in \arg\min_{\mathbf{u}_t\in
\bar{\mathcal{C}}_t^1\times\cdots\times\bar{\mathcal{C}}_t^N}
J_t(\mathbf{u}_t)
\label{eq:global_one_point_optimization_short}
\end{equation}
is the collective packet-selection problem. Therefore, if each local problem has a unique minimizer, then the global problem also has a unique minimizer. In summary, the receiver-side packet-selection problem is a finite exact optimization that is globally defined but rigorously separable across receivers. Each receiver selects the packet that minimizes its local overlap-consistency cost, and the collection of these local minimizers simultaneously solves the collective problem.

\section{Simulation}
\label{sec:sim}

\pgfplotstableread[col sep=comma]{sim_data/baseline_history.csv}\histtab
\pgfplotstableread[col sep=comma]{sim_data/comm_cost.csv}\commtab
\pgfplotstableread[col sep=comma]{sim_data/cost_history.csv}\costtab
\pgfplotstablecreatecol[create col/expr accum={\pgfmathaccuma+\thisrow{t_sender}}{0}]{cs}\costtab
\pgfplotstablecreatecol[create col/expr accum={\pgfmathaccuma+\thisrow{t_receiver}}{0}]{cr}\costtab
\pgfplotstablecreatecol[create col/expr accum={\pgfmathaccuma+\thisrow{n_admitted}}{0}]{ca}\costtab
\pgfplotstablegetrowsof{\histtab}
\pgfmathtruncatemacro{\LR}{\pgfplotsretval-1}
\pgfplotstablegetrowsof{\costtab}
\pgfmathtruncatemacro{\CR}{\pgfplotsretval-1}
\newcommand{\Hc}[1]{\pgfplotstablegetelem{\LR}{#1}\of\histtab
  \pgfmathprintnumber[fixed,fixed zerofill,precision=2]{\pgfplotsretval}}
\newcommand{\Hp}[1]{\pgfplotstablegetelem{\LR}{#1}\of\histtab
  \pgfmathparse{100*\pgfplotsretval}%
  \pgfmathprintnumber[fixed,fixed zerofill,precision=2]{\pgfmathresult}}
\newcommand{\Cn}[3]{\pgfplotstablegetelem{#1}{#2}\of\commtab
  \pgfmathprintnumber[fixed,fixed zerofill,precision=#3]{\pgfplotsretval}}
\newcommand{\Kc}[2]{\pgfplotstablegetelem{\CR}{#1}\of\costtab
  \pgfmathprintnumber[fixed,fixed zerofill,precision=#2]{\pgfplotsretval}}
\newcommand{\Eta}[2]{%
  \pgfplotstablegetelem{\LR}{nrmse_self}\of\histtab\edef\tmpa{\pgfplotsretval}%
  \pgfplotstablegetelem{\LR}{#2}\of\histtab\edef\tmpb{\pgfplotsretval}%
  \pgfplotstablegetelem{#1}{scalars}\of\commtab\edef\tmpc{\pgfplotsretval}%
  \begingroup\pgfkeys{/pgf/fpu=true,/pgf/fpu/output format=fixed}%
  \pgfmathparse{\tmpc/((\tmpa-\tmpb)*1000000)}%
  \global\let\EtaVal\pgfmathresult\endgroup
  \pgfmathprintnumber[fixed,fixed zerofill,precision=2]{\EtaVal}}

\subsection{Setup}
The IPD-GP method is evaluated on a decentralized mapping task over a compact domain $\mathcal{X}=[-6,6]^2$, discretized on a $50\times 50$ grid. The collection of agents consists of $N=4$ mobile sensors. As shown in Fig.~\ref{fig:truth}, each agent $i$ is assigned an axis-aligned square subdomain of half-width $3.6$ centered at $(\pm 2.4,\pm 2.4)$. The four subdomains collectively cover $\mathcal{X}$, and every pair overlaps, forming axis-aligned shared geometric regions. 

The communication graph is fully connected, yielding 12 directed edges. Over $T=50$ clock steps, every agent draws one noisy measurement from its own subdomain. All agents share homogeneous hyperparameters to align with the underlying assumptions of the compared aggregation baselines, length scale $\ell=1.50$, signal standard deviation $\sigma_f=0.70$, and noise standard deviation $\sigma_n=0.09$. Under the IPD-GP protocol, each sender selects $p=10$ inducing points per directed edge per step, and each receiver admits at most one packet per step. 

Performance is evaluated using four metrics, defined as in
\cite{muzaffar.Kontoudis.Stilwell2026}.  The normalized root-mean-square error
(NRMSE) is the root-mean-square prediction error divided by the range of $f$
over $\mathcal{X}$, and the negative log predictive density (NLPD) is the mean
negative log Gaussian predictive density at the evaluation points, taken with
the observation noise variance included, so that it scores the predicted
variance together with the mean.  Both are pooled over the combined
subdomains.  Overlap RMSE (oRMSE) and overlap NLPD (oNLPD) restrict the same
two scores to the shared strips, where the agents interact.

\subsection{Results and Baselines}
The IPD-GP is compared against four estimators: (1) Self-only, where each agent regresses solely on its own measurements without exchange; (2) DEC-gPoE and (3) DEC-grBCM, which are decentralized aggregation rules implemented via discrete-time average consensus using 50 iterations per grid node; and (4) Centralized GP, providing the all-data upper bound.

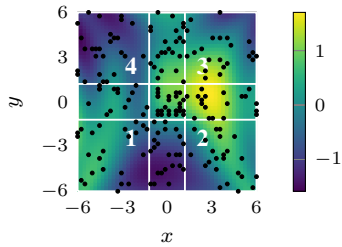
\begin{figure}[t]
\centering
\begin{tikzpicture}
\begin{axis}[
  colormap/viridis,
  width=0.53\columnwidth, axis equal image,
  xmin=-6,xmax=6,ymin=-6,ymax=6,
  xtick={-6,-3,0,3,6}, ytick={-6,-3,0,3,6},
  xlabel={$x$}, ylabel={$y$},
  tick label style={font=\scriptsize}, label style={font=\footnotesize},
  colorbar, colorbar style={width=0.16cm,tick label style={font=\scriptsize}}]
\addplot[matrix plot*,point meta=explicit,mesh/cols=50,mesh/ordering=y varies]
  table[x=x,y=y,meta=z,col sep=comma]{sim_data/truth_grid.csv};
\addplot[white,no marks,line width=0.7pt] coordinates {(-6,-6) (1.2,-6) (1.2,1.2) (-6,1.2) (-6,-6)};
\addplot[white,no marks,line width=0.7pt] coordinates {(-1.2,-6) (6,-6) (6,1.2) (-1.2,1.2) (-1.2,-6)};
\addplot[white,no marks,line width=0.7pt] coordinates {(-1.2,-1.2) (6,-1.2) (6,6) (-1.2,6) (-1.2,-1.2)};
\addplot[white,no marks,line width=0.7pt] coordinates {(-6,-1.2) (1.2,-1.2) (1.2,6) (-6,6) (-6,-1.2)};
\addplot[only marks,mark=*,mark size=0.7pt,black] table[x=x,y=y,col sep=comma]{sim_data/measurements_agent_1.csv};
\addplot[only marks,mark=*,mark size=0.7pt,black] table[x=x,y=y,col sep=comma]{sim_data/measurements_agent_2.csv};
\addplot[only marks,mark=*,mark size=0.7pt,black] table[x=x,y=y,col sep=comma]{sim_data/measurements_agent_3.csv};
\addplot[only marks,mark=*,mark size=0.7pt,black] table[x=x,y=y,col sep=comma]{sim_data/measurements_agent_4.csv};
\node[white,font=\bfseries\small] at (axis cs:-2.4,-2.4) {1};
\node[white,font=\bfseries\small] at (axis cs: 2.4,-2.4) {2};
\node[white,font=\bfseries\small] at (axis cs: 2.4, 2.4) {3};
\node[white,font=\bfseries\small] at (axis cs:-2.4, 2.4) {4};
\end{axis}
\end{tikzpicture}
\vspace{-0.3cm}
\caption{Scalar field with four square white agent subdomains and $T=50$ sampling locations (black dots).}
\label{fig:truth}
\vspace{-0.3cm}
\end{figure}

\begin{table}[t]
\centering
\caption{Accuracy over the combined subdomains at $t=T$ and associated communication costs. $^{\dagger}$Total number of scalars transmitted over the run. $^{\ddagger}$Millions of scalars per unit of NRMSE reduction relative to Self-only.}
\label{tab:main}
\setlength{\tabcolsep}{2pt}
\scriptsize

\begin{tabular}{l cc cc c r r}
\toprule
& \multicolumn{2}{c}{$\times 10^{-2}$} & & & & & \\
\cmidrule(lr){2-3}
Method
& NRMSE
& oRMSE
& NLPD
& oNLPD
& Cov.
& Scalars$^{\dagger}$
& $\eta^{\ddagger}$ \\
\midrule

IPD-GP
& \Hp{nrmse_prop}
& \Hp{ormse_prop}
& \Hc{nlpd_prop}
& \Hc{onlpd_prop}
& \Hc{cov_prop}
& \Cn{0}{scalars}{0}
& \Eta{0}{nrmse_prop} \\

Self-only
& \Hp{nrmse_self}
& \Hp{ormse_self}
& \Hc{nlpd_self}
& \Hc{onlpd_self}
& \Hc{cov_self}
& 0
& --- \\

DEC-gPoE
& \Hp{nrmse_gpoe}
& \Hp{ormse_gpoe}
& \Hc{nlpd_gpoe}
& \Hc{onlpd_gpoe}
& \Hc{cov_gpoe}
& \Cn{1}{scalars}{0}
& \Eta{1}{nrmse_gpoe} \\

DEC-grBCM
& \Hp{nrmse_grbcm}
& \Hp{ormse_grbcm}
& \Hc{nlpd_grbcm}
& \Hc{onlpd_grbcm}
& \Hc{cov_grbcm}
& \Cn{2}{scalars}{0}
& \Eta{2}{nrmse_grbcm} \\

Centralized GP
& \Hp{nrmse_cen}
& \Hp{ormse_cen}
& \Hc{nlpd_cen}
& \Hc{onlpd_cen}
& \Hc{cov_cen}
& ---
& --- \\

\bottomrule
\end{tabular}
\vspace{-0.6cm}
\end{table}

\begin{figure*}[t]
\centering
\pgfplotsset{
  bsl/.style={width=3.3cm,height=2.9cm,scale only axis,
    xmin=1,xmax=50,xtick={1,25,50},
    title style={font=\footnotesize,yshift=-4pt},
    tick label style={font=\scriptsize}, label style={font=\footnotesize},
    every axis plot/.append style={line width=0.8pt},
    xlabel={step $t$},
    grid=both,grid style={gray!18,line width=0.25pt}}}
\begin{tikzpicture}
\begin{groupplot}[group style={group size=4 by 1,horizontal sep=1.05cm},bsl]

\nextgroupplot[title={NRMSE},ymode=log,log ticks with fixed point,
  ytick={0.03,0.06,0.12,0.24},
  legend to name=bsllegend,legend columns=5,
  legend style={font=\footnotesize,draw=none,
    /tikz/every even column/.append style={column sep=10pt}}]
\addplot[blue]                 table[col sep=comma,x=t,y=nrmse_prop]  {sim_data/baseline_history.csv};
\addlegendentry{IPD-GP}
\addplot[gray,dashed]          table[col sep=comma,x=t,y=nrmse_self]  {sim_data/baseline_history.csv};
\addlegendentry{Self-only}
\addplot[orange]               table[col sep=comma,x=t,y=nrmse_gpoe]  {sim_data/baseline_history.csv};
\addlegendentry{DEC-gPoE}
\addplot[red]                  table[col sep=comma,x=t,y=nrmse_grbcm] {sim_data/baseline_history.csv};
\addlegendentry{DEC-grBCM}
\addplot[black,densely dotted] table[col sep=comma,x=t,y=nrmse_cen]   {sim_data/baseline_history.csv};
\addlegendentry{Centralized GP}

\nextgroupplot[title={Overlap RMSE},ymode=log,log ticks with fixed point,
  ytick={0.08,0.16,0.32,0.64}]
\addplot[blue]                 table[col sep=comma,x=t,y=ormse_prop]  {sim_data/baseline_history.csv};
\addplot[gray,dashed]          table[col sep=comma,x=t,y=ormse_self]  {sim_data/baseline_history.csv};
\addplot[orange]               table[col sep=comma,x=t,y=ormse_gpoe]  {sim_data/baseline_history.csv};
\addplot[red]                  table[col sep=comma,x=t,y=ormse_grbcm] {sim_data/baseline_history.csv};
\addplot[black,densely dotted] table[col sep=comma,x=t,y=ormse_cen]   {sim_data/baseline_history.csv};

\nextgroupplot[title={NLPD}]
\addplot[blue]                 table[col sep=comma,x=t,y=nlpd_prop]   {sim_data/baseline_history.csv};
\addplot[gray,dashed]          table[col sep=comma,x=t,y=nlpd_self]   {sim_data/baseline_history.csv};
\addplot[orange]               table[col sep=comma,x=t,y=nlpd_gpoe]   {sim_data/baseline_history.csv};
\addplot[red]                  table[col sep=comma,x=t,y=nlpd_grbcm]  {sim_data/baseline_history.csv};
\addplot[black,densely dotted] table[col sep=comma,x=t,y=nlpd_cen]    {sim_data/baseline_history.csv};

\nextgroupplot[title={Overlap NLPD}]
\addplot[blue]                 table[col sep=comma,x=t,y=onlpd_prop]  {sim_data/baseline_history.csv};
\addplot[gray,dashed]          table[col sep=comma,x=t,y=onlpd_self]  {sim_data/baseline_history.csv};
\addplot[orange]               table[col sep=comma,x=t,y=onlpd_gpoe]  {sim_data/baseline_history.csv};
\addplot[red]                  table[col sep=comma,x=t,y=onlpd_grbcm] {sim_data/baseline_history.csv};
\addplot[black,densely dotted] table[col sep=comma,x=t,y=onlpd_cen]   {sim_data/baseline_history.csv};
\end{groupplot}
\end{tikzpicture}

\vspace{1pt}
\pgfplotslegendfromname{bsllegend}
\vspace{-0.35cm}
\caption{Accuracy over the combined subdomains against time.  The first two
panels use a logarithmic ordinate, and panels two and four restrict the score
to the shared strips.}
\label{fig:baseline}
\vspace{-0.6cm}
\end{figure*}
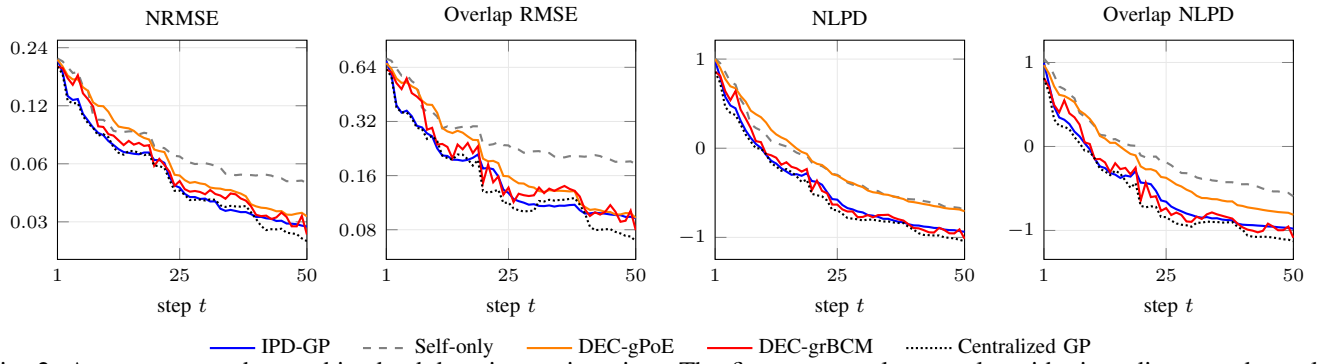

\section{Discussion}
\label{sec:discussion}

As detailed in Table~\ref{tab:main}, the communication cost of IPD-GP is light, requiring only $0.024$ million scalars to be transmitted over the entire run because IPD-GP senders transmit a small, targeted batch of inducing points and their posterior moments. In contrast, DEC-gPoE and DEC-grBCM require $3.0$ and $4.5$ million scalars, respectively, because they rely on iterative consensus broadcast across every query point on the grid. 

Furthermore, the computation cost of the IPD-GP approach is approximately half that of the BCM approaches. The execution speed of IPD-GP is faster due to the fact that much less data exists to be processed, and far fewer rounds of calculation are required to assimilate the compact packets compared to running dense average consensus at every spatial node. 

The predictive accuracy of IPD-GP is close to what the aggregation baselines reach.
Over the combined subdomains it sits between DEC-gPoE and DEC-grBCM, and the
gap to the centralized GP reference is small.  DEC-grBCM is the more accurate
of the two decentralized rules, but it reaches that accuracy at more than
$150\times$ the transmitted scalars per unit of error reduction against
self-only regression.  These results offer evidence that overlap-targeted
packet exchange with persistent assimilation recovers much of the accuracy of
consensus-based aggregation at a fraction of its communication.  They rest on a single realization, so repeated trials over independent field and measurement
draws are needed before the margins can be stated with confidence.

\section{Conclusion}
\label{sec:conclusion}
An Inducing Points Decentralized Gaussian Process (IPD-GP) framework is developed for scalable, multi-agent scalar-field mapping. By exploiting spatial discrepancies across overlapping subdomains, agents exchange compact, geometry-targeted data packets rather than computing globally fused summaries. The developed method utilizes a receiver-specific transmission rule to select
inducing points over shared boundaries, coupled with a receiver-side admission
gate.  Assimilating any packet lowers the receiver's modeled variance, so the
gate does not test whether uncertainty falls but whether the sender reports a
sharper posterior at that location, which raises the chance that the reduction
reflects better information rather than the conditional-independence model
under which it is computed. Simulation results confirm that IPD-GP reduces overlap estimation error and matches the predictive accuracy of data-heavy aggregation methods like DEC-grBCM. This accuracy is reached while transmitting two orders of magnitude fewer scalars than either aggregation rule.

Future research will expand this framework across several interconnected areas. A primary focus involves establishing a formal theoretical foundation for the IPD-GP architecture, including convergence guarantees for the persistent pseudo-measurement recursion and analytical error bounds for assimilating correlated neighbor packets. The current variance-based admission rule will be extended into a comprehensive trust mechanism resilient to untrustworthy or Byzantine agents that might broadcast corrupted statistical moments.

\bibliographystyle{IEEETrans.bst}
\small
\bibliography{scc,sccmaster,muzaffar,Hossein_LCSS}

\begin{thebibliography}{10}
\def\url#1{}
\csname url@samestyle\endcsname
\providecommand{\newblock}{\relax}
\providecommand{\bibinfo}[2]{#2}
\providecommand{\BIBentrySTDinterwordspacing}{\spaceskip=0pt\relax}
\providecommand{\BIBentryALTinterwordstretchfactor}{4}
\providecommand{\BIBentryALTinterwordspacing}{\spaceskip=\fontdimen2\font plus
\BIBentryALTinterwordstretchfactor\fontdimen3\font minus
  \fontdimen4\font\relax}
\providecommand{\BIBforeignlanguage}[2]{{%
\expandafter\ifx\csname l@#1\endcsname\relax
\typeout{** WARNING: IEEEtran.bst: No hyphenation pattern has been}%
\typeout{** loaded for the language `#1'. Using the pattern for}%
\typeout{** the default language instead.}%
\else
\language=\csname l@#1\endcsname
\fi
#2}}
\providecommand{\BIBdecl}{\relax}
\BIBdecl

\bibitem{SCC.Sabella.1998}
P.~Sabella, ``A rendering algorithm for visualizing 3{D} scalar fields,''
  \emph{ACM SIGGRAPH Computer Graphics}, vol.~22, no.~4, pp. 51--58, 1998.

\bibitem{SCC.Lin.Liu.ea2019}
\BIBentryALTinterwordspacing
Z.~Lin, H.~H.~T. Liu, and M.~Wotton, ``Kalman filter based large-scale wildfire
  monitoring with a system of {{UAV}s},'' \emph{IEEE Trans. Ind. Electron.},
  vol.~66, no.~1, pp. 606--615, Jan. 2019.
  \url{https://ieeexplore.ieee.org/document/8331879}
\BIBentrySTDinterwordspacing

\bibitem{SCC.Ren.Li.ea2021}
\BIBentryALTinterwordspacing
T.~Ren, H.~Li, M.~F. Modest, and C.~Zhao, ``Efficient two-dimensional scalar
  fields reconstruction of laminar flames from infrared hyperspectral
  measurements with a machine learning approach,'' \emph{J. Quant. Spectrosc.
  Radiat. Transfer}, vol. 271, p. 107724, 2021.
  \url{https://www.sciencedirect.com/science/article/pii/S002240732100217X}
\BIBentrySTDinterwordspacing

\bibitem{SCC.Rasmussen.Williams2006}
C.~E. Rasmussen and C.~K.~I. Williams, \emph{{G}aussian processes for machine
  learning.}\hskip 1em plus 0.5em minus 0.4em\relax Cambridge, MA: MIT Press,
  2006.

\bibitem{SCC.Qureshi.Ogri.ea2024}
\BIBentryALTinterwordspacing
M.~Qureshi, T.~E. Ogri, Z.~I. Bell, and R.~Kamalapurkar, ``Scalar field mapping
  with adaptive high-intensity region avoidance,'' in \emph{Proc. IEEE Conf.
  Control Technol. Appl.}, Newcastle upon Tyne, UK, Aug. 2024, pp. 388--393.
  \url{https://ieeexplore.ieee.org/document/10666509}
\BIBentrySTDinterwordspacing

\bibitem{Lederer2019Uniform}
A.~Lederer, J.~Umlauft, and S.~Hirche, ``Uniform error bounds for {G}aussian
  process regression with application to safe control,'' in \emph{Advances in
  Neural Information Processing Systems}, vol.~32, 2019, pp. 659--669.

\bibitem{Ma2018Sampling}
K.-C. Ma, L.~Liu, H.~K. Heidarsson, and G.~S. Sukhatme, ``Data-driven learning
  and planning for environmental sampling,'' \emph{J. Field Robot.}, vol.~35,
  no.~5, pp. 643--661, 2018.

\bibitem{Suryan2020MinTime}
V.~Suryan and P.~Tokekar, ``Learning a spatial field in minimum time with a
  team of robots,'' \emph{IEEE Trans. Robot.}, vol.~36, no.~5, pp. 1562--1576,
  2020.

\bibitem{Santos2021Coverage}
M.~Santos, U.~Madhushani, A.~Benevento, and N.~E. Leonard, ``Multi-robot
  learning and coverage of unknown spatial fields,'' in \emph{Proc. Int. Symp.
  Multi-Robot Multi-Agent Syst. (MRS)}, 2021, pp. 137--145.

\bibitem{Jang2020ActiveSensing}
D.~Jang, J.~Yoo, C.~Y. Son, D.~Kim, and H.~J. Kim, ``Multi-robot active sensing
  and environmental model learning with distributed {G}aussian process,''
  \emph{IEEE Robot. Autom. Lett.}, vol.~5, no.~4, pp. 5905--5912, 2020.

\bibitem{Brunke2022SafeLearning}
L.~Brunke, M.~Greeff, A.~W. Hall, Z.~Yuan, S.~Zhou, J.~Panerati, and A.~P.
  Schoellig, ``Safe learning in robotics: From learning-based control to safe
  reinforcement learning,'' \emph{Annual Review of Control, Robotics, and
  Autonomous Systems}, vol.~5, pp. 411--444, 2022.

\bibitem{Liu2020Survey}
H.~Liu, Y.-S. Ong, X.~Shen, and J.~Cai, ``When {G}aussian process meets big
  data: A review of scalable {GPs},'' \emph{IEEE Trans. Neural Netw. Learn.
  Syst.}, vol.~31, no.~11, pp. 4405--4423, 2020.

\bibitem{Gielis2022Critical}
J.~Gielis, A.~Shankar, and A.~Prorok, ``A critical review of communications in
  multi-robot systems,'' \emph{Current Robot. Rep.}, vol.~3, no.~4, pp.
  213--225, 2022.

\bibitem{SCC.Ogri.Qureshi.ea2025a}
T.~E. Ogri, M.~Qureshi, Z.~I. Bell, W.~A. Makumi, and R.~Kamalapurkar, ``An
  output feedback approach to differential graphical games in networked linear
  systems,'' \emph{IEEE Control Syst. Lett.}, vol.~9, pp. 2265--2270, Sep.
  2025.

\bibitem{qureshi2025gaussianprocessbasedscalarfield}
M.~Qureshi, T.~E. Ogri, H.~Ramos, Z.~I. Bell, and R.~Kamalapurkar, ``Gaussian
  process-based scalar field estimation in {GPS-}denied environments,'' 2025.

\bibitem{Cortes2009Kriged}
J.~Cort{\'e}s, ``Distributed kriged {K}alman filter for spatial estimation,''
  \emph{IEEE Trans. Autom. Control}, vol.~54, no.~12, pp. 2816--2827, 2009.

\bibitem{Park2018Patchwork}
C.~Park and D.~Apley, ``Patchwork {K}riging for large-scale {G}aussian process
  regression,'' \emph{J. Mach. Learn. Res.}, vol.~19, no.~7, pp. 1--43, 2018.

\bibitem{Hinton2002PoE}
G.~E. Hinton, ``Training products of experts by minimizing contrastive
  divergence,'' \emph{Neural Comput.}, vol.~14, no.~8, pp. 1771--1800, 2002.

\bibitem{Cao2014gPoE}
Y.~Cao and D.~J. Fleet, ``Generalized product of experts for automatic and
  principled fusion of {G}aussian process predictions,''
  \emph{arXiv:1410.7827}, 2014.

\bibitem{Tresp2000BCM}
V.~Tresp, ``A {B}ayesian committee machine,'' \emph{Neural Comput.}, vol.~12,
  no.~11, pp. 2719--2741, 2000.

\bibitem{Deisenroth2015Distributed}
M.~P. Deisenroth and J.~W. Ng, ``Distributed {G}aussian processes,'' in
  \emph{Proc. Int. Conf. Mach. Learn. (ICML)}, 2015, pp. 1481--1490.

\bibitem{liu2018generalizedrobustbayesiancommittee}
\BIBentryALTinterwordspacing
H.~Liu, J.~Cai, Y.~Wang, and Y.-S. Ong, ``Generalized robust bayesian committee
  machine for large-scale gaussian process regression,'' 2018.
  \url{https://arxiv.org/abs/1806.00720}
\BIBentrySTDinterwordspacing

\bibitem{Rulliere2018Nested}
D.~Rulli{\`e}re, N.~Durrande, F.~Bachoc, and C.~Chevalier, ``Nested {K}riging
  predictions for datasets with a large number of observations,''
  \emph{Statist. Comput.}, vol.~28, no.~4, pp. 849--867, 2018.

\bibitem{Chen2015Fusion}
J.~Chen, K.~H. Low, Y.~Yao, and P.~Jaillet, ``{G}aussian process decentralized
  data fusion and active sensing for spatiotemporal traffic modeling and
  prediction in mobility-on-demand systems,'' \emph{IEEE Trans. Autom. Sci.
  Eng.}, vol.~12, no.~3, pp. 901--921, 2015.

\bibitem{Pillonetto2019Finite}
G.~Pillonetto, L.~Schenato, and D.~Varagnolo, ``Distributed multi-agent
  {G}aussian regression via finite-dimensional approximations,'' \emph{IEEE
  Trans. Pattern Anal. Mach. Intell.}, vol.~41, no.~9, pp. 2098--2111, 2019.

\bibitem{Kontoudis2021Nested}
G.~P. Kontoudis and D.~J. Stilwell, ``Decentralized nested {G}aussian processes
  for multi-robot systems,'' in \emph{Proc. IEEE Int. Conf. Robot. Autom.
  (ICRA)}, 2021, pp. 8881--8887.

\bibitem{Yuan2024Lightweight}
Z.~Yuan and M.~Zhu, ``Lightweight distributed {G}aussian process regression for
  online machine learning,'' \emph{IEEE Trans. Autom. Control}, vol.~69, no.~6,
  pp. 3928--3943, 2024.

\bibitem{muzaffar.Kontoudis.Stilwell2026}
G.~P. Kontoudis and D.~J. Stilwell, ``Multi-agent federated learning using
  covariance-based nearest neighbor {G}aussian processes,'' \emph{IEEE Trans.
  Mach. Learn. Commun. Netw.}, vol.~4, pp. 115--138, 2026.

\bibitem{Kontoudis2024Federated}
------, ``Scalable, federated {G}aussian process training for decentralized
  multi-agent systems,'' \emph{IEEE Access}, vol.~12, pp. 77\,800--77\,815,
  2024.

\bibitem{Xie2019ADMM}
A.~Xie, F.~Yin, Y.~Xu, B.~Ai, T.~Chen, and S.~Cui, ``Distributed {G}aussian
  processes hyperparameter optimization for big data using proximal {ADMM},''
  \emph{IEEE Signal Process. Lett.}, vol.~26, no.~8, pp. 1197--1201, 2019.

\bibitem{Yue2024FederatedGP}
X.~Yue and R.~Kontar, ``Federated {G}aussian process: Convergence, automatic
  personalization and multi-fidelity modeling,'' \emph{IEEE Trans. Pattern
  Anal. Mach. Intell.}, vol.~46, no.~6, pp. 4246--4261, 2024.

\bibitem{QuinoneroCandela2005Unifying}
J.~Qui{\~n}onero-Candela and C.~E. Rasmussen, ``A unifying view of sparse
  approximate {G}aussian process regression,'' \emph{J. Mach. Learn. Res.},
  vol.~6, no.~65, pp. 1939--1959, 2005.

\bibitem{Snelson2005Pseudo}
E.~Snelson and Z.~Ghahramani, ``Sparse {G}aussian processes using
  pseudo-inputs,'' in \emph{Adv. Neural Inf. Process. Syst.}, vol.~18, 2005,
  pp. 1257--1264.

\bibitem{Titsias2009Variational}
M.~Titsias, ``Variational learning of inducing variables in sparse {G}aussian
  processes,'' in \emph{Proc. 12th Int. Conf. Artif. Intell. Statist.}, 2009,
  pp. 567--574.

\bibitem{Datta2016NNGP}
A.~Datta, S.~Banerjee, A.~O. Finley, and A.~E. Gelfand, ``Hierarchical
  nearest-neighbor {G}aussian process models for large geostatistical
  datasets,'' \emph{J. Amer. Stat. Assoc.}, vol. 111, no. 514, pp. 800--812,
  2016.

\bibitem{Gramacy2015LocalGP}
R.~B. Gramacy and D.~W. Apley, ``Local {G}aussian process approximation for
  large computer experiments,'' \emph{J. Comput. Graph. Statist.}, vol.~24,
  no.~2, pp. 561--578, 2015.

\bibitem{Cole2021LIGP}
D.~A. Cole, R.~B. Christianson, and R.~B. Gramacy, ``Locally induced gaussian
  processes for large-scale simulation experiments,'' \emph{Statistics and
  Computing}, vol.~31, no.~3, p.~33, 2021.

\bibitem{Hoang2019Collective}
T.~N. Hoang, Q.~M. Hoang, K.~H. Low, and J.~P. How, ``Collective online
  learning of {G}aussian processes in massive multi-agent systems,'' in
  \emph{Proc. AAAI Conf. Artif. Intell.}, vol.~33, no.~1, 2019, pp. 7850--7857.

\bibitem{Lederer2021RealTime}
A.~Lederer, A.~J. Ord\'o\~nez Conejo, K.~Maier, W.~Xiao, J.~Umlauft, and
  S.~Hirche, ``{G}aussian process-based real-time learning for safety critical
  applications,'' in \emph{Proceedings of the 38th International Conference on
  Machine Learning}, ser. PMLR, vol. 139, 2021.

\end{thebibliography}

\end{document}